\documentstyle[12pt]{article}
\def\beq{\begin{equation}}
\def\eeq{\end{equation}}
\def\bea{\begin{eqnarray}}
\def\eea{\end{eqnarray}}
\def\bq{\begin{quote}}
\def\eq{\end{quote}}

\def\PLB{{\it Phys. Lett.} }

\def\NP{{\it Nucl.Phys.} }
\def\PR{{\it Phys.Rev.} }

\def\gappeq{\mathrel{\rlap {\raise.5ex\hbox{$>$}}
{\lower.5ex\hbox{$\sim$}}}}

\def\lappeq{\mathrel{\rlap{\raise.5ex\hbox{$<$}}
{\lower.5ex\hbox{$\sim$}}}}

\begin{document}
\topmargin -0.5cm
\oddsidemargin -0.3cm
\evensidemargin -0.8cm
\pagestyle{empty}
\vspace*{5mm}
\begin{center}
{\bf ESTIMATES OF THE $O(\alpha^{4}_{s})$ CORRECTIONS} \\
{\bf TO $\sigma_{tot} (e^{+}e^{-} \to {\rm hadrons})$,
$\Gamma (\tau \to \nu_{\tau} + {\rm hadrons})$ } \\
{\bf AND DEEP INELASTIC SCATTERING SUM RULES} \\
\vspace*{1cm}
{\bf Andrei L. Kataev} \footnote{Permanent address: Institute for
Nuclear Research of the Russian Academy of Sciences, 117312 Moscow,
Russia} \\
\vspace{0.3cm}
Theoretical Physics Division, CERN, \\
CH-1211 Geneva 23, Switzerland \\
\vspace{0.5cm}
and \\
\vspace*{0.5cm}
{\bf Valery V. Starshenko} \\
\vspace*{0.3cm}
330091 Zaporozhye, Ukraine \\
\vspace*{2cm}
{\bf ABSTRACT} \\
\end{center}
\vspace*{5mm}
\noindent
We present the estimates of the order $O(\alpha^{4}_{s})$ QCD
corrections to
$R(s) = \sigma_{tot} (e^{+}e^{-} \to {\rm hadrons}) /\\
\sigma (e^{+}e^{-} \to \mu^{+} \mu^{-}),
R_{\tau} = \Gamma (\tau \to \nu_{\tau} + {\rm hadrons}) /
\Gamma (\tau \to \nu_{\tau} \overline{\nu}_{e} e)$
and to the deep inelastic scattering sum rules, namely to the
non-polarized and polarized Bjorken sum rules and to the Gross--Llewellyn
Smith sum rule. The estimates are obtained in the $\overline{MS}$-scheme
using the principle of minimal sensitivity and the
effective charges approach.
\vspace*{2cm}



\newpage
\setcounter{page}{1}
\pagestyle{plain}
\section{Introduction}

During the last few years, essential progress has been achieved in the
area of the calculation of the next-next-to-leading order (NNLO) QCD
corrections to the number of physical quantities.  Indeed, the
complete NNLO $O(\alpha^{3}_{s})$ QCD corrections are known at present
for the characteristics of $e^{+} e^{-} \to \mbox{hadrons}$ process
\cite{aaa}, \cite{bb}, $\tau \to \nu_{\tau} +
\mbox{hadrons}$ decay \cite{cc}, \cite{dd} and for the deep inelastic
scattering sum rules, namely the non-polarized Bjorken sum rule (BjnSR)
\cite{ee}, the Gross-Llewellyn Smith sum rule (GLSSR) and the
polarized Bjorken sum rule (BjpSR) \cite{ff}.  Amongst the physical
information provided by these results is the estimate of the
theoretical uncertainties of the corresponding next-to-leading-order
(NLO) perturbative QCD predictions for
$R(s) = \sigma_{tot}(e^{+} e^{-} \to {\rm hadrons}) /
\sigma (e^{+} e^{-} \to \mu^{+} \mu^{-})$ \cite{ggg},
$\Gamma (Z^{0} \to {\rm hadrons})$ (see. e.g., \cite{hh}),
$R_{\tau} = \Gamma (\tau \to \nu_{\tau} + {\rm hadrons}) /
\Gamma (\tau \to \nu_{\tau} \overline{\nu}_{e} e)$ \cite{cc},
BjnSR \cite{jj}, GLSSR and BjpSR \cite{kk}.

In view of the fact that the precision of the experimental data for
$\sigma_{tot} (e^{+} e^{-} \to {\rm hadrons}),\\
\Gamma (Z^{0} \to {\rm hadrons})$ and for the structure functions
of the deep inelastic scattering is continuously increasing, the
problem of estimating the effects of the
next-after-next-next-to-leading order (NANNLO) $O(\alpha^{3}_{s})$
corrections to the measurable physical quantities now arises.

The first attempt to build the bridge between the results of the
concrete calculations of the order $O(\alpha^{3}_{s})$ corrections to
$R(s)$ and the higher-order coefficients of the corresponding
perturbative QCD series was made in Ref. \cite{lll}. However, it was
later demonstrated that the considerations of Ref. \cite{lll} have
definite drawbacks \cite{mm}, \cite{nn}.  A certain step in the
direction of more substantiated estimates of the order
$O(\alpha^{4}_{s})$ QCD corrections was made in the case of $R_{\tau}$
in Ref. \cite{cc}.  These estimates are based on the tendency,
observed in Ref. \cite{oo}, of the scheme-dependent uncertainties of
the perturbative QCD predictions for $R(s)$ and $R_{\tau}$ to decrease
as a result of taking into account the order $O(\alpha^{3}_{s})$-terms.
The foundations of Ref. \cite{oo} were further generalized in the
process of phenomenological studies of the QCD predictions for
$R_{\tau}$, taking into account the concrete pieces of the higher-order
perturbative QCD corrections \cite{pp} (for the earlier development of
a similar technique, see Ref. \cite{qq}).

It should be stressed that the analysis of Ref. \cite{cc} does not
allow us to fix the sign of the order $O(\alpha^{4}_{s})$ contribution
to $R_{\tau}$.  In this work we generalize the QED ideas of Refs.
\cite{rr}, \cite{ss} and \cite{tt} to the QCD case and estimate the
order $O(\alpha^{3}_{s})$ and $O(\alpha^{4}_{s})$ corrections to
$R(s)$, $R_{\tau}$ and deep inelastic scattering sum rules with the
help of the principle of minimal sensitivity (PMS) \cite{rr} and the
effective charges (ECH) approach \cite{uu}, which is equivalent {\it a
posteriori} to the scheme-invariant perturbation theory \cite{vv}.
Contrary to the conclusions of Ref. \cite{ww}, we argue that the
application of these approaches allows one to control the value and fix
the sign of the higher-order QCD corrections to physical quantities,
once the preceding ones are known in the particular scheme.

\section{The Description of the Formalism}

Consider first the order $O(a^{N})$ approximation of a physical quantity
\beq
D_{N} = d_{0} a(1 + \sum^{N-1}_{i-1} \, d_{i} a^{i})
\label{1}
\eeq
with $a = \alpha_{s}/\pi$ being the solution of the corresponding
renormalization group equation for the $\beta$-function which is
defined as
\beq
\mu^{2} \frac{\partial a}{\partial \mu^{2}} =
\beta (a) = - \beta_{0} a^{2}
(1 +  \sum^{N-1}_{i-1} \, c_{i} a^{i})\ .
\label{2}
\eeq
In the process of the concrete calculations of the coefficients
$d_{i}, i \geq 1$ and $c_{i}, i \geq 2$, the $\overline{MS}$ scheme is
commonly used.  However, this scheme is not the unique prescription for
fixing the RS ambiguities.  In both phenomenological and theoretical
studies other methods are also widely applied.

The PMS \cite{rr} and ECH \cite{uu} prescriptions stand out from these
other methods.  Indeed, they are based on the conceptions of the
scheme-invariant quantities, which are defined as the combinations of
the scheme-dependent coefficients in Eqs. (1) and (2).  Both these
methods can claim to be the ``optimal" prescriptions, in the sense that
they provide better convergence of the corresponding approximations in
the non-asymptotic regime, and thus allow an estimation of the
uncertainties of the perturbative series
in the definite order of perturbation theory.  Therefore,
the applications of the ``optimal" methods allow one to estimate the
effects of the order $O(a^{N+1})$-corrections starting from the
approximations $D^{opt}_{N} (a_{opt})$ calculated in a certain
``optimal" approach \cite{rr} - \cite{tt}.  This idea is closely
related to the QED technique of Ref. \cite{zz}, which was used to
predict  the renormalization-group controllable $ln(m_{\mu}/m_e)$-terms
in the series for
$(g-2)_{\mu}$ from the expression of $(g-2)_{\mu}$ through the
effective coupling constant $\bar{\alpha} (m_{\mu} / m_{e})$. In our
work we are using a similar technique to estimate the constant terms of
the higher-order corrections in QCD. \footnote{The application of this
similar technique to the analysis of the five-loop approximations of
$(g-2)_{\mu}$ and $(g-2)_{e}$ will be considered elsewhere.}

Let us re-expand $D_{N}^{opt} (a_{opt})$ in terms of the coupling
constant $a$ of the particular scheme
\beq
D_{N}^{opt} (a_{opt}) = D_{N} (a) + \delta D_{N}^{opt} a^{N+1}
\label{3}
\eeq
where
\beq
\delta D_{N}^{opt} = \Omega_{N}(d_{i}, c_{i}) -
\Omega_{N} (d_{i}^{opt}, c_{i}^{opt})
\label{4}
\eeq
are the  numbers which simulate the coefficients of the
order $O(a^{N+1})$-corrections to the physical quantity, calculated in
the particular initial scheme, say the $\overline{MS}$-scheme.  The
coefficients $\Omega_{N}$ can be obtained from the following system of
equations:
\bea
\frac{\partial}{\partial \tau}
(D_{N} + \Omega_{N} a^{N+1}) =
O(a^{N+2}), \nonumber \\
\frac{\partial}{\partial c_{i}}
(D_{N} + \Omega_{N} a^{N+1}) =
O(a^{N+2}),\ i \geq 2
\label{5}
\eea
where the parameter $\tau = \beta_{0} \ell n (\mu^2 / \Lambda^2)$
represents the freedom in the choice of the renormalization point
$\mu$. The conventional scale parameter $\Lambda$ will not explicitly
appear in all our final formulas.

The explicit form of the coefficients $\Omega_{2}$ and $\Omega_{3}$ in
which we will be interested can be obtained by the solution of the
system of equations (5), following the lines of Ref. \cite{rr}.  We
present here only the final expressions:
\beq
\Omega_{2} = d_{0}d_{1} (c_{1} + d_{1}),
\label{6}
\eeq
\beq
\Omega_{3} = d_{0}d_{1} (c_{2} + \frac{1}{2} c_{1}d_{1}
-2d_{1}^{2} + 3d_{2})\ .
\label{7}
\eeq

It should be stressed that in the ECH approach $d_{i}^{ECH} \equiv 0$
for all $i \geq 2$.  Therefore one gets the following expressions for
the NNLO and NANNLO corrections in Eq. (3):
\beq
\delta D_{2}^{ECH} = \Omega_{2} (d_{1}, c_{1})
\label{8}
\eeq
\beq
\delta D_{2}^{ECH} = \Omega_{3}(d_{1}, d_{2}, c_{1}, c_{2})
\label{9}
\eeq
where $\Omega_{2}$ and $\Omega_{3}$ are defined in Eqs. (6) and (7)
respectively.

In order to find similar corrections to Eq. (3) in the N-th order of
perturbation theory starting from the PMS approach \cite{rr}, it is
necessary to use the relations obtained in Ref. \cite{yy} between the
coefficients $r_{i}^{PMS}$ and $c_{i}^{PMS} \; (i \geq 1)$ in the
expression for the order $O(a^{N}_{PMS})$ approximation
$D^{PMS}_{N} (a_{PMS})$ of the physical quantity under consideration.
The corresponding corrections have the following form:
\beq
\delta D_{2}^{PMS} = \delta D_{2}^{ECH} - \frac{d_{0} c^{2}_{1}}{4}
\label{10}
\eeq
\beq
\delta D_{3}^{PMS} = \delta D_{3}^{ECH}\ .
\label{11}
\eeq
Notice the identical coincidence of the NANNLO corrections obtained
starting from both the PMS and ECH approaches.  A similar observation
was made in Ref. \cite{ss} using different (but related) considerations.

\section{Input of the Analysis}

Consider first the familiar characteristic of the $e^{+}e^{-} \to
\gamma \to {\rm hadrons}$ process, namely the $D$-function defined in
the Euclidean region:
\beq
D(Q^{2}) = Q^{2} \int^{\infty}_{0} \, \frac{R(s)}{(s+Q^{2})^{2}} \, ds
\label{12}
\eeq
Its perturbative expansion has the following form:
\bea
D(Q^{2}) & = & 3 \Sigma Q^{2}_{f} [1 + a + d_{1}a^{2} + d_{2}a^{3} +
d_{3}a^{4} + ... ]  \nonumber \\
& + & (\Sigma Q_{f})^{2} [ \tilde{d}_{2} a^{3} + O(a^{4}) ]
\label{13}
\eea
where $Q_{f}$ are the quark charges, and the structure proportional to
$(\Sigma Q_{f})^{2}$ comes from the light-by-light-type diagrams. The
coefficients $d_{1}$ and $d_{2}, \tilde{d}_{2}$ were calculated in the
$\overline{MS}$-scheme in Refs. \cite{ggg} and \cite{aaa} respectively.
They have the following numerical form:
\bea
d_{1}^{\overline{MS}} & \approx & 1.986 - 0.115 f \nonumber \\
d_{2}^{\overline{MS}} & \approx & 18.244 - 4.216 f + 0.086 f^{2} ,\
\tilde{d}_{2} \approx -1.240\ .
\label{14}
\eea

Following the proposals of Ref. \cite{aai}, we will treat the
light-by-light-type term in Eq. (13) separately from the ``main"
structure of the $D$-function, which is proportional to the
quark-parton expression $D^{QP}(Q^{2}) = 3 \Sigma Q^{2}_{f}$. In fact,
one can hardly expect that it is possible to predict higher-order
coefficients $\tilde{d}_{i}, i \geq 3$ of the second structure in Eq.
(13) using the only explicitly-known term $\tilde{d}_{2}$.  Therefore
we will neglect the light-by-light-type structure as a whole in all our
further considerations.  This approximation is supported by the
relatively tiny contribution of the second structure of Eq. (13) to the
final NNLO correction to the $D$-function.

The next important ingredient of our analysis is the QCD
$\beta$-function (2), which is known in the MS-like schemes at the NNLO
level \cite{bbi}.  Its corresponding coefficients read
\bea
\beta_{0} & = & (11 - \frac{2}{3} f) \frac{1}{4} \approx
2.75 - 0.167 f \nonumber \\
c_{1} & = & \frac{153 - 19f}{66 - 4 f} \nonumber \\
c_{2}^{\overline{MS}} & = & \frac{77139 - 15099f + 325 f^{2}}
{9504 - 576 f}\ .
\label{16}
\eea
Using now the perturbative expression for the $D$-function, one can
obtain the perturbtive expression for $R(s)$, namely
\bea
R(s) & = & 3 \Sigma Q^{2}_{f} [1 + a_{s} + r_{1} a^{2}_{s} +
r_{2} a^{3}_{s} + r_{3} a^{4}_{s} + ... ] \nonumber \\
& + & (\Sigma Q_{f})^{2} [ \tilde{r}_{2} a^{3}_{s} + ... ]
\label{17}
\eea
where $a_{s} = \bar{\alpha}_{s} / \pi$, and
\bea
r_{1} & = & d_{1} \nonumber \\
r_{2} & = & d_{2} - \frac{\pi^{2} \beta^{2}_{0}}{3}, \,\
\tilde{r}_{2} = \tilde{d}_{2} \nonumber \\
r_{3} & = & d_{3} - \pi^{2} \beta^{2}_{0}
(d_{1} + \frac{5}{6} c_{1})\ .
\label{18}
\eea
The corresponding $\pi^{2}$-terms come from the analytic continuation
of the Euclidean result for the $D$-function to the physical region.
The effects of the higher-order $\pi^{2}$-terms were discussed in
detail in Ref. \cite{cci}.

The perturbative expression for $R_{\tau}$ is defined as
\bea
R_{\tau} & = & 2 \int_{0}^{M^{2}_{\tau}} \, \frac{ds}{M^{2}_{\tau}} \,
(1 - s/M^{2}_{\tau})^{2} \, (1 + 2s/M^{2}_{\tau}) \tilde{R}(s)
\nonumber \\
& \simeq & 3[1 + a_{\tau} + r_{1}^{\tau} a^{2}_{\tau} +
r_{2}^{\tau} a^{3}_{\tau} + r_{3}^{\tau} a^{4}_{\tau} + ...]
\label{19}
\eea
where $a_{\tau} = \alpha_{s} (M^{2}_{\tau}) / \pi$ and $\tilde{R} (s)$
is $R(s)$ with
with $f = 3, (\Sigma Q_{f})^{2} = 0, 3 \Sigma Q^{2}_{f}$ substituted
for $3 \Sigma \mid V_{ff'} \mid^{2}$ and $\mid V_{ud} \mid^{2} +
\mid V_{us} \mid^{2} \approx 1$.

It can be shown that in the
$\overline{MS}$-scheme the coefficients of the
series (18) are related to those of the series (13) for the
$D$-function by the following numerical relations (see Refs. \cite{cc},
\cite{qq}):
\bea
r^{\tau}_{1} & = &
d_{1}^{\overline{MS}} (f = 3) + 3.563 = 5.202 \nonumber \\
r^{\tau}_{2} & = &
d_{2}^{\overline{MS}} (f = 3) + 19.99 = 26.366 \nonumber
\\
r^{\tau}_{3} & = &
d_{3}^{\overline{MS}} (f = 3) + 78.00\ .
\label{20}
\eea
In order to estimate the values of the order $O(a^{3})$ and $O(a^{4})$
corrections to $R(s)$ and $R_{\tau}$, we will apply Eqs. (6) - (11) in
the Euclidean region to the perturbative series for the $D$-function
and then obtain the corresponding estimates using Eqs. (17) and (19).
This is one of the origins of the disagreement of our considerations
with those presented in Ref. \cite{ww}.

We now recall the perturbative expression for the non-polarized Bjorken
deep-inelastic scattering sum rule
\bea
BjnSR & = & \int^{1}_{0} \, F_{1}^{\bar{v}p - vp} (x, Q^{2}) dx
\nonumber \\
& = & 1 - \frac{2}{3} a (1 + d_{1}a + d_{2}a^{2} + d_{3} a^{3} + ...)
\label{21}
\eea
where the coefficients $d_{1}$ and $d_{2}$ are known in the
$\overline{MS}$
scheme from the results of calculations of Ref. \cite{jj} and Ref.
\cite{ee} respectively:
\beq
d_{1}^{\overline{MS}} \approx 5.75 - 0.444 f
\label{22}
\eeq
\beq
d_{2}^{\overline{MS}} \approx 54.232 - 9.497 f + 0.239 f^{2}\ .
\label{23}
\eeq
The expression for the polarized Bjorken sum rule BjpSR has the
following form:
\bea
BjpSR & = & \int^{1}_{0} \, g_{1}^{ep-en} (x, Q^{2}) dx \nonumber \\
& = & \frac{1}{3} \mid \frac{g_A}{g_V} \mid \,
\{ 1-a (1 + d_{1}a + d_{2} a^{2} + d_{3} a^{3} + ... ) \}
\label{24}
\eea
where the coefficients $d_{1}$ and $d_{2}$ were explicitly calculated
in the $\overline{MS}$ scheme in
Refs. \cite{kk} and \cite{ff} respectively.
The results of these calculations read
\beq
d_{1}^{\overline{MS}} \approx 4.583 - 0.333 f
\label{25}
\eeq
\beq
d_{2}^{\overline{MS}} \approx 41.440 - 7.607 f + 0.177 f^{2}\ .
\label{26}
\eeq
It should be stressed that since deep inelastic scattering sum rules
are defined in the Euclidean region, we can directly apply to them the
methods discussed in Section 2, and thus predict the values of the
coefficients $d_{3}$ using Eqs. (9) and (7) without any additional
modifications.

It is also worth emphasizing that, in spite of the identical
coincidence of the NLO correction to the Gross-Llewellyn Smith sum rule
\bea
GLSSR & = & \frac{1}{2}\int^{1}_{0} \, F_3^{\overline{\nu}p+\nu p}
(x,Q^2)dx \nonumber \\
& = & 3\{1-a(1+d_{1}a+d_{2}a^{2}+d_{3}a^{3}+...)\}
\label{gls}
\eea
with the result of Eq. (24) \cite{kk}, the corresponding NNLO correction
differs from the result of Eq. (25) by the contributions of the
light-by-light-type terms typical of the GLSSR \cite{ff}:
\bea
(d_{2})_{GLSSR}  =  (d_{2})_{BjpSR}+ \tilde{d}_{2} ,\
\tilde{d}_{2}  =  -0.413f
\label{lbl}
\eea
Since these light-by-light-type
terms appear for the first time at the NNLO, it is impossible to
predict the value of the light-by-light-type contribution at the NANNLO
level using the corresponding NNLO terms as the input information.
However, noticing that at the NNLO level
the corresponding light-by-light-type contributions are small,
we will assume that the similar contributions
are small at the NANNLO level also. After this assumption
our estimates of the NNLO and NANNLO corrections to the BjpSR can be
considered also as the estimates of the corresponding
corrections in the perturbative series for th GLSSR . Note, that
the Pad\'e predictions of the order $O(a^{4})$ contributions
to the GLSSR \cite{ddi}, which do not take into account the necessity
of the careful considerations of the light-by-light-type terms, have
definite drawbacks.  However, certain other results of Refs. \cite{ddi}
and \cite{sli} deserve comparison with the results of our studies.

\section{Outputs of the Analysis}

The estimates of the coefficients of the order $O(a^{3})$ and
$O(a^{4})$ QCD corrections to the $D$-function, $R(s)$, BjnSR and
BjpSR/GLSSR obtained following
the discussions of Section 2 with the help of
the results summarized in Section 3, are presented in Tables 1 - 4
respectively.  Due to the complicated $f$-dependence of the
coefficients $\Omega_{2}, \Omega_{3}$ in Eqs. (6) and (7), we are
unable to predict the explicit $f$-dependence of the corresponding
coefficients in the form respected by perturbation theory.  The
results are presented for the fixed number of quark flavours $1 \leq f
\leq 6$ and are related to the $\overline{MS}$ scheme.
The estimates of
the NNLO corrections, obtained starting from both the ECH and PMS
approaches, are compared with the results of the explicit calculations.

Using the results of Table 1 and Eqs. (19), we are able to predict the
NNLO coefficient of $R_{\tau}$:
\beq
\left ( r^{\tau}_{2} \right )^{est}_{ECH} \approx 25.6
\label{27}
\eeq
\beq
\left ( r^{\tau}_{2} \right )^{est}_{PMS} \approx 24.8\ .
\label{28}
\eeq
One can see the agreement with the explicitly calculated result
\beq
\left ( r^{\tau}_{3} \right )_{\overline{MS}} = 26.366\ .
\label{29}
\eeq
Considering this agreement as the additional {\it a posteriori} support
of the methods used, we use the estimate of the NANNLO coefficient for
the $D$-function with $f = 3$ numbers of flavours as presented in Table
1, namely:
\beq
d^{est}_{3} = 27.46
\label{30}
\eeq
and predict the value of the NANNLO coefficient of $R_{\tau}$:
\beq
\left (r^{\tau}_{3} \right )^{est} \approx 105.5
\label{31}
\eeq

\section{Discussions}

We are now ready to discuss the main results of our analysis.
\begin{enumerate}
\item
The predictions of the NNLO corrections to the $D$-function, BjnSR and
BjpSR/GLSSR are in qualitative agreement
with the results of the explicit
calculations of Refs. \cite{aaa} \cite{ee} and \cite{ff} and respect
the tendency of the corresponding coefficients to decrease with
increasing number of flavours.
\item
The best agreement of the NNLO estimates with the exact results is
obtained for the case $f = 3$.  This fact, which we do not understand,
supports the application of the method used for estimating the NNLO and
NANNLO corrections to $R_{\tau}$.
\item
The positive feature of the estimates as obtained by us is their
scheme-dependence.  Notice that since the methods used correctly
reproduce the renormalization-group-controllable terms \cite{zz},
\cite{rr}, the transformation from the $\overline{MS}$ scheme to other
variants of the MS-like scheme will not spoil the qualitative agreement
with the results of the explicit calculations.
\item
The results of Table 2 demonstrate that the $\pi^{2}$ effects give
dominating contributions to the coefficients of $R(s)$.
\item
{}From the phenomenological point of view, the most interesting results
are the estimates of the NANNLO corrections to $R_{\tau}$ (see Eq.
(32)) and $R(s)$  for $f = 5$ numbers of flavours.  Taking $\alpha_{s}
(M_{Z}) \approx 0.12$, we get the estimate of the corresponding NANNLO
contribution to both $\Gamma (Z^{0} \to {\rm hadrons})$ and
$\Gamma (Z^{0} \to \bar{q}q)$:
\beq
\delta \Gamma_{Z^{0}} \approx -97 (a(M_{Z}))^{4} \approx -2 \times
10^{-4}\ .
\label{32}
\eeq
It is of the order of magnitude of other corrections included in the
current analysis of LEP data (see, e.g., \cite{eei}).
\item
Taking $\alpha_{s}(M_{\tau}) \approx 0.36$ \cite{cc}, we get the
numerical estimate of the NANNLO contribution to $R_{\tau}$:
\beq
\delta R_{\tau} \approx 105.5 \, a^{4}_{\tau}
\approx 1.8 \times 10^{-2}\ .
\label{33}
\eeq
It is larger than the recently calculated power-suppressed perturbative
\cite{ffi} and non-perturbative \cite{ggi}
contributions to $R_{\tau}$.
\item
Our estimate for $d_{3} (f = 3)$ of Eq. (31) is more definite than the
one presented in Ref. \cite{sli}, namely $d_{3} (f = 3) =
55^{+60}_{-24}$, and than the bold guess estimate $d_{3} (f = 3) = \pm
25$, given in Ref. \cite{pp}.  Moreover, the related estimate of the
NANNLO contribution to $R_{\tau}$ (see Eq. (32)) is more precise than
those presented in Refs. \cite{cc} and \cite{pp}, namely $\delta
R_{\tau} = \pm 130 \, a^{4}_{\tau}$ \cite{cc} and $\delta R_{\tau} =
(78 \pm 25) a^{4}_{\tau}$ \cite{pp}, \footnote{A similar, more
conservative estimate gives $\delta R_{\tau} = (78 \pm 50)
a^{4}_{\tau}$ (for a review, see Ref. \cite{hhi}).} and is smaller than
the result of applying the Pad\'e resummation technique directly to
$R_{\tau}$, namely $\delta R_{\tau} = 133 a^{4}_{\tau}$ \cite{sli}.
\item
The qualitative agreement of the results of Tables 3 and 4 for BjnSR
and BjpSR with the corresponding Pad\'e estimates of Ref. \cite{ddi}
can be considered as the argument in favour of the applicability of
both theoretical methods in the Euclidean region for the concrete
physical applications. Let us stress again that in the process of
these applications the light-by-light-type structures, contributing
say to $R(s)$ and GLSSR, should be treated seperately.

\item
Notice also   , that the application of the Pad\'e resummation
technique to $R(s)$  \cite{sli}, stimulated by the previous similar
studies of Ref. \cite{jji}, gives less definite estimates than the
results of applications of our methods (compare the estimate $\delta
R(s) = \left( -49^{+54}_{-40} \right ) a^{4}$ \cite{sli} with the
result $\delta R(s) = -97 a^{4}$ from Table 2 and Eq. (33)).  This fact
is related with the problems of applicability of the Pad\'e resummation
technique to the sign variation perturbative series, obtained from the
explicit NNLO approximation of the $D$-function with $f = 5$ numbers of
flavours \cite{aaa} (see also Ref. \cite{bb}).
\end{enumerate}

\section{Acknowledgements}

We wish to thank K.H. Becks, D. Yu. Bardin and S. Bethke for
stimulating the presentation of the results of this work prior to
publication, at the III Workshop on Artificial Intelligence and Expert
Systems in HEP (Oberammergau, October 1993), LEP Meetings at CERN
(March 1994) and the University of Aachen (April 1994) respectively.

We are grateful to R.N. Faustov for attracting our attention to the
detailed consideration of the results of Ref. \cite{ss} at the
preliminary stage  of our similar QED studies, which will be presented
elsewhere and to F. Le Diberder for useful comments.

The work of one of us (V.V.S.) was supported during his stay in Moscow
by the Russian Fund for Fundamental Research, Grant No. 93-02-14428.

{\bf Note added.} Our estimate for $d_3$ was recently supported by the
phenomenological analysis of the ALEPH data for $R_{\tau}$ \cite{Dib}.

\newpage
\begin{center}
\begin{tabular}{|c|c|c|c|c|}
 \hline
$f$ & $d^{ex}_{2}$ & $(d^{est}_{2})_{ECH}$ & $(d^{est}_{2})_{PMS}$ &
$d^{est}_{3}$ \\ \hline
1 & 14.11 & 7.54 & 7.70 & 75.4 \\ \hline
2 & 10.16 & 6.57 & 7.55 & 49.76 \\  \hline
3 & 6.37 & 5.61 & 6.40 & 27.46 \\ \hline
4 & 2.76 & 4.68 & 5.27 & 8.37 \\  \hline
5 & -0.69 & 3.77 & 4.16 & -7.7 \\ \hline
6 & -3.96 & 2.88 & 3.08& -20.89 \\ \hline
\end{tabular}
\end{center}
Table 1 : The results of estimates of the NNLO and NANNLO corrections in
the series for the
$D$-functions.\\
\begin{center}
\begin{tabular}{|c|c|c|c|c|} \hline
$f$ & $r^{ex}_{2}$ & $(r^{est}_{2})_{ECH}$ & $(r^{est}_{2})_{PMS}$ &
$r^{est}_{3}$ \\ \hline
1 & -7.84 & -14.41 & -13.16 & -166.4 \\ \hline
2 & -9.04 & -12.63 & -11.65 & -146.6 \\ \hline
3 & -10.27 & -11.03 & -10.23 & -128.4 \\ \hline
4 & -11.52 & -9.58 & -8.98 & -111.8 \\ \hline
5 & -12.76 & -8.29 & -7.9 & -96.8 \\ \hline
6 & -14.01 & -7.17 & -6.97 & -83.3 \\ \hline
\end{tabular}
\end{center}
Table 2: The results of estimates of the NNLO and NANNLO corrections in
the series for $R(s)$.\\
\begin{center}
\begin{tabular}{|c|c|c|c|c|} \hline
$f$ & $d^{ex}_{2}$ & $(d^{est}_{2})_{ECH}$ & $(d^{est}_{2})_{PMS}$ &
$d^{est}_{3}$\\ \hline
1 & 44.97 & 39.62 & 40.78 & 423.77 \\ \hline
2 & 36.19 & 33.28 & 34.26& 302.7 \\ \hline
3 & 27.89 & 27.37 & 28.16 & 199.61 \\ \hline
4 & 20.07 & 21.91 & 22.50 & 113.68 \\ \hline
5 & 12.72 & 16.91 & 17.30 & 44.1 \\ \hline
6 & 5.85 & 12.39 & 12.59 & -9.89 \\ \hline
\end{tabular}
\end{center}
Table 3: The results of estimates of the NNLO and NANNLO corrections in
the series for BjnSR.\\
\begin{center}
\begin{tabular}{|c|c|c|c|c|} \hline
$f$ & $d^{ex}_{2}$ & $(d^{est}_{2})_{ECH}$ & $(d^{est}_{2})_{PMS}$ &
$d^{est}_{3}$ \\ \hline
1 & 34.01 & 27.25 & 28.41 & 290.27 \\ \hline
2 & 26.93 & 23.11 & 24.09 & 203.37 \\ \hline
3 & 20.21 & 19.22 & 20.01 & 129.96 \\ \hline
4 & 13.84 & 15.57 & 16.16 & 68.07 \\ \hline
5 & 7.83 & 12.19 & 12.59& 17.79 \\ \hline
6 & 2.17 & 9.08 & 9.29 & -21.57 \\ \hline
\end{tabular}
\end{center}
Table 4: The results of estimates of the NNLO and NANNLO corrections in
the series for BjpSR and GLSSR.\\

\end{document}